\documentclass[journal=jacsat,manuscript=article]{achemso}

\usepackage{chemformula} 
\usepackage[T1]{fontenc} 

\author{V. J. Ajith}
\affiliation[Indian Institute of Science Education and Research Pune]
{Department of Physics, Indian Institute of Science Education and Research Pune, Pune-411008, Maharashtra, India}
\author{Shivprasad Patil}
\affiliation[Indian Institute of Science Education and Research Pune]
{Department of Physics, Indian Institute of Science Education and Research Pune, Pune-411008, Maharashtra, India}
\email{s.patil@iiserpune.ac.in}

\title[Translational diffusion of a fluorescent tracer molecule in nanoconfined water]
{Translational diffusion of a fluorescent tracer molecule in nanoconfined water}

\begin{document}
	
	\begin{tocentry}
		\begin{center}
  
    	\includegraphics[width=\textwidth]{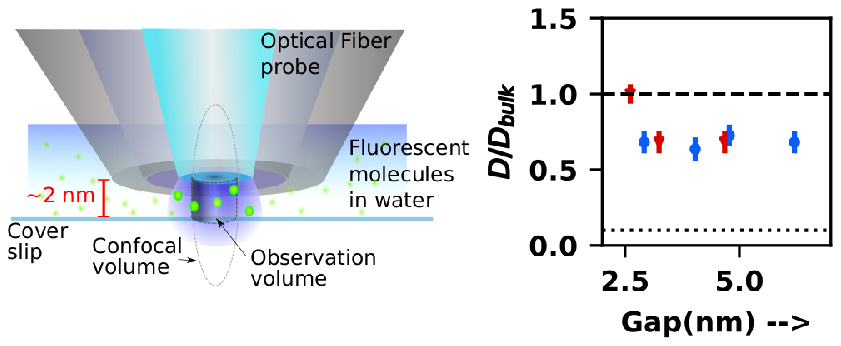}

        \end{center}		
	\end{tocentry}
	
	\begin{abstract}
		Diffusion of tracer dye molecules in water confined to nanoscale  is an important subject  with a direct bearing on many technological applications. It is not yet clear however,  if the dynamics of water in hydrophilic as well as hydrophobic nanochannels remains bulk-like. Here, we present diffusion measurement of a fluorescent dye molecule in water confined to nanoscale between two hydrophilic surfaces whose separation can be controlled with a precision of less than a nm. We observe that the  fluorescence intensities correlate over a fast($\sim$ 30 $\mu$s) and slow ($\sim$ 1000 $\mu$s) time components. The slow timescale is due to adsorption of fluorophores to the confining walls and it disappears in presence of 1 M salt. The fast component is attributed to  diffusion of dye molecules in the gap and is found to be bulk-like for sub-10 nm separations  and indicates that viscosity of water under confinement remains unaltered up to confinement gap as small as $\sim$ 5 nm. Our findings contradict some of the recent measurements of diffusion under nanoconfinement, however they are consistent with many estimates of self-diffusion using  molecular dynamics simulations and measurements using neutron scattering experiments.
	\end{abstract}
	
\section{\label{intro}Introduction}

Water is a an important solvent responsible for macromolecular   organization, which is vital for life on earth\cite{lfinney_overview_1996}.  For the most part, the water involved in these reactions is interfacial, or tightly confined to  nanoscale cavities. Besides the  scientific importance, investigation of nanoconfined water is useful in many applications such as  filtration and desalination \cite{das_carbon_2014}, fuel cells \cite{buchi_investigation_2001}, properties of clays, minerals and food materials\cite{aines_water_1984}. It also plays an important role in biological processes of  protein folding \cite{levy_water_2006} and  transport in cellular pores \cite{takata_aquaporins_2004}.

In these situations,   water is  confined between either  hydrophobic or hydrophilic  solid walls, which are roughly separated by a few nm.  Such nanoconfined water, through many experiments, is known to exhibit curious response to external perturbations such as mechanical shear or electrical fields.     
Among these are  fast permeation of water in carbon nanotubes \cite{majumder_enhanced_2005}, permeation of water through helium-leak-tight membrane \cite{nair_unimpeded_2012}, dynamic solidification of water layers next to the surface \cite{khan_dynamic_2010}, suppression of freezing \cite{alabarse_freezing_2012}, and anomalously low out-of-plane dielectric constant\cite{fumagalli_anomalously_2018}. Through decades of research, there is some  understanding of physical origins of these curious properties of water at nanoscale. However, the subject is active and not free from controversies\cite{li_nonlinear_2008, raviv_fluidity_2001, zhu_viscosity_2001,khan_nacl-dependent_2016} 
Many spectroscopic measurements have been  performed  with the purpose of shedding light on behavior of  water residing at interfaces,  nanoconfined  between walls or contained in nanoscale pores\cite{morishige_x-ray_1997,dokter_anomalous_2005}.

In particular, viscosity  or the damping coefficient  of confined water is measured using Surface Force Apparatus \cite{raviv_fluidity_2001}, Atomic Force Microscope\cite{jeffery_direct_2004}, and Capillary action\cite{vchurayev_measurement_1970}. Few dissipation measurements have  claimed that confined water has a viscosity similar to bulk\cite{raviv_fluidity_2001}, while others concluded 
an 8-order increase in viscosity\cite{li_nonlinear_2008}. Nanoscale water also exhibits rheological response with characteristic timescales of the order of 1 to 10 $\mu$s and  analogies are drawn with  slow-down in supercooled water\cite{li_nonlinear_2008}. This means that,  self diffusion of water molecules or tracer diffusion of dye molecule through it,  is expected to show signatures of dynamic heterogeneity inherent to glass-forming systems. Hence,  diffusion measurements become central to answer the question- How does water exhibit slow response to both electrical and mechanical perturbations? .

The slow-down in dynamics, however,   is not an universal feature of all spectroscopic measurements of diffusion so far.   For instance,   measurements  using Quasi Elastic Neutron Scattering(QENS) in room temperature \cite{bellissent-funel_single-particle_1995,takahara_neutron_2005}  conclude that the self translational diffusion of nanconfined water in porous materials is same as bulk water within an order. While at low temperatures below bulk-freezing point, nanoconfined water remains fluid and has almost 100 times  reduction in translational diffusion\cite{weigler_static_2020,luisespehr_dynamics_2011}.
 
There are few measurements of tracer diffusion in water confined to  single nanochannels. Using Fluorescence Correlation Spectroscopy(FCS) \cite{de_santo_subdiffusive_2010,gravelle_flow-induced_2019},  it was reported  that there is no slow-down in diffusion of  tracer dye. However, by tracking the fluorescence distribution along nanochannels over time, it was shown that  diffusion is slowed down by factor of 100 in organic solvents\cite{zhong_fluorescence_2018}. These measurements are relevant to the newly  emerging field of nanofluidics\cite{eijkel_nanofluidics_2005,bocquet_nanofluidics_2010,sparreboom_principles_2009}. It has applications in iontronics\cite{karnik_electrostatic_2005,yan_nanofluidic_2009}, energy harvesting\cite{ren_slip-enhanced_2008} and sequencing of biopolymers\cite{marie_integrated_2013,chinappi_protein_2018}. There are few reports of subdiffusion of proteins in nanochannels\cite{de_santo_subdiffusive_2010}. In this context it is important to know about the factors affecting the dynamics of solute molecules in the solution confined in nanochannels. The interfacial charge, viscosity and confinement effects are expected to have an effect, however it is still unclear beyond what  length scale  we can ignore interfacial effects.

Over last few decades, the questions regarding  dynamics of water confined between hydrophilic walls separated by less than 1 nm are addressed through experiments\cite{khan_dynamic_2010,li_nonlinear_2008,zhu_viscosity_2001}. This question is also tractable using molecular dynamics simulation and large body of work exists in literature\cite{leng_fluidity_2005,sendner_interfacial_2009}. However, there is a lack of clarity on properties of water where confining walls are separated by distance in the range of 1 -10 nm, particularly about the  dynamics of solute molecules. This is of immense interest to the newly emerging field of nanofluidics which has tremendous technological impact.    

Here, we report diffusion of tracer dye,  Coumarin (Cu343) in water confined to sub-10 nm separations using an instrument developed in our laboratory. An Aluminium-coated fiber tip, with an optical opening at its end is used to confine water between itself and the flat glass substrate. Using a servo control the gap between the tip and substrate is controlled with precision of less than a nm.  The instrument allows us to probe diffusion of tracer dyes in the nanoscale gap which can be controlled in-situ. We clearly observed two timescales for diffusing dye indicating  co-existence of slow and fast regions for diffusion of the dye.

The disappearance of slow  component after addition of 1 M salt, indicates that it  is likely due to fluorophores bound to the glass surface. Further, we show that the diffusion coefficient is bulk-like for sub-10 nm confinement gaps.  The slow-down in tracer diffusion seen in the past  can be attributed to the adsorption-desorption to the confining walls.

\begin{figure}
	\includegraphics[width=0.98\textwidth]{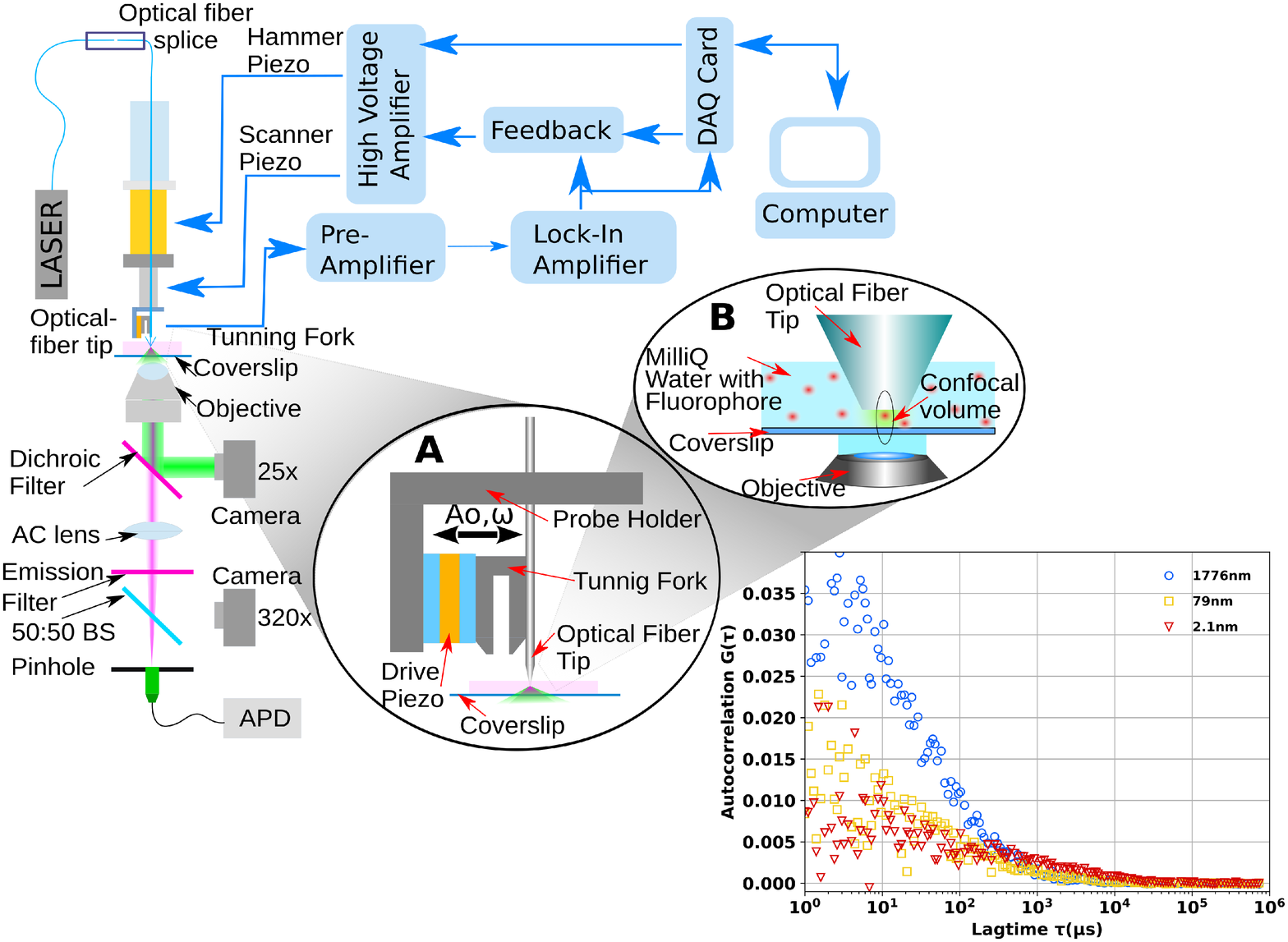}
	\caption{A schematic of the instrument used to perform measurements. A sharp tip  made out of single mode optical fiber coated with Aluminium is loaded onto one prong of a quartz crystal tuning fork. A small optical aperture is opened at the end using focused ion beam. A servo control is used to keep the tip  at a fixed separation from the substrate  using amplitude of the tuning fork prong as input to the feedback controller.  The excitation light from laser is coupled into the optical fiber and reaches the gap between the tip and substrate. The emitted light from the tracer dye is collected using an objective lens and is fed into an Avalanche Photodiode. Representative autocorrelation curves of fluctuations in fluorescence intensities from  Coumarin 343 dye diffusing in the  gap between the tip and substrate held at three fixed separations ( 2 nm, 79 nm and 1776 nm) shown.  The lateral amplitude of the tip is used as an input to  active feedback to control the separation. The data at large separations is collected without the feedback control. All separations are determined using voltages provided to the scanner piezo and   have systematic errors of 2 nm as mentioned in the text.\label{fig1}    }
\end{figure}	

\section{Experiments}
All experiments are performed  with a home-built instrument\cite{ajith_new_2020}. A schematic of the instrument is shown in figure \ref{fig1}.   In order to create a small controlled gap  in which water is nano-confined, a sharp tip is pulled out of single mode optical fiber is used(Thorlabs 460HP, 125 $\mu m$ diameter with cladding). The water is confined between the tip and the substrate. The tip is coated with 150 nm of aluminium and before mounting on one prong  of Quartz Crystal Tunig Fork ( QTF), an optical aperture (200-900 nm diameter) is opened at its end using Focused Ion Beam(FIB) milling. This constitutes one  of  the two flat confining surfaces.  The other end of the fiber is used to couple the light from a 446 nm continuous wave laser with  40 mW power. Before coupling, the power of the laser is reduced to 1 mW using neutral density filters. After coupling, around 5 to 50 $\mu W$ of light comes out of the tip. The QTF is then mounted on a piezo drive fixed to the probe holder (Figure \ref{fig1}A). 
The liquid cell is fitted with a circular cover-slip which forms the other flat confining surface. This bottom confining surface is cleaned using UV-Ozone treatment.  The cell is  filled with pure water (MilliQ, $~$18 M$\Omega$ cm) having 25-50 nM concentration of tracer dye in it. The QTF is mechanically oscillated on its resonance ($\sim$ 32 kHz) using the piezo drive and amplitude of the prong with tip is   $<$ 1 nm. This oscillation generates a piezo electric current of same frequency and  its  amplitude  monotonically decreases as the tip-coverslip gap decreases. Before every experiment the QTF current amplitude is recorded with lock-in amplifier as a function of piezo extension. In the experiments we monitor and record the QTF current amplitude and use it to calculate the gap. In a typical experiment,  the tip is approached towards the substrate with servo-control,  in which the QTF current amplitude serves as input parameter to control the separation between the tip and the substrate. This defines the thickness of the water confined between  two walls and can be controlled using an active feedback. Since we use 16- bit resolution DAQ for our analog outputs for piezo extension, the relative position of the tip can be controlled with precision of 0.1 nm. For gaps typically above 10-15 nm, the QTF current is not sensitive to the gap changes, so we cannot use the feedback. For such large separations, we give constant voltage to the piezo which corresponds to a specific gap. All  separations are ‘average’ distances and are  not directly measured in our experiments. They are calculated from the voltages provided to the scanner piezos.

The special optical fiber tip is capable of illuminating a local volume in the nanoconfined water beneath the tip. The diffusing fluorophores enter the nanoconfined water and transit through the  illuminated volume to give fluorescence. A confocal arrangement is used to collect these fluorescence photons and an Avalanche Photodiode(APD) is used to detect them. Typically,  the tip diameter is larger than the cross section of confocal volume at the beam waist. It ensures that only the fluorescence from the nanoconfined region is detected.  The observation volume is the convolution of confocal volume and the nanoconfined gap produced by the tip and substrate.  
It is  a  disc-like volume with height of tip-sample gap and diameter of the confocal volume.  The number of fluorophores in the observation volume fluctuates due to diffusion. For low concentrations, these fluctuations will be seen as the fluctuations in the detected fluorescence intensity. These fluctuations are correlated for the average time the fluorophore spends in observation volume. This average residential timescale can be extracted by taking the autocorrelation of the fluctuations in the fluorescence intensity.  If the diffusion under confinement is significantly different from the bulk,  the slow-down should give increase in this residential timescale. If there is a fraction of fluorophores undergoing adsorption-desorption events at the interface along with diffusion then those fluorophores spend more time in observation volume. In this case the autocorrelation will have another slow residential time scale along with the residential time scale of the freely diffusing molecules. 

Each autocorrelation of fluorescent intensity fluctuations for a specific gap is calculated for 100 seconds using a correlator card(Correlator Flex99-12). The correlator card uses multiple-tau algorithm to calculate the autocorrelation\cite{wohland_standard_2001} and samples the time trace with a precision of 12.5 ns.

Using the instrument we performed experiments with fluorescent Coumarin 343 molecules(11-oxo-2,3,5,6,7,11-Hexahydro-1H-pyrano[2,3-f]pyrido[3,2,1-ij]quinoline-10-carboxylic acid;\ch{C16H15NO4}). In experiments,  fluorescent molecules are mixed in  water to get $~$25-50 nM concentration by series dilution. For a certain set of  experiments fluorescent molecule is mixed in 1M NaCl water solution to  avoid adsorption of  tracer dye on the glass surface.    
The bulk diffusion coefficient of Coumarin 343  is 550 $nm^2/\mu s$(5.5 $\times$ 10$^{-10} m^2/s$)\cite{sasmal_diffusion_2011}. 
As per the 3D visualising tool "Jmol"\cite{Jmol}, maximum length along x,y,z dimensions of Coumarin 343 is ($\sim$ 1.15 nm, $\sim$ 0.7 nm, $\sim$ 0.3 nm ). 

The uncertainty in  the gap between the tip and coverslip in  our instrument is $\sim$ 2 nm. This is mainly determined by roughness in the tip and roughness of coverslip for the area beneath the tip. Using a commercial AFM we measured the roughness of the coverslip. For an area of 1$\mu$m x 1$\mu$m, coverslip has a RMS roughness of 0.5$\pm$0.2nm. The optical fiber tips are sliced with Galium ions (Ga+) which are  accelerated by 30 kV and current of 40pA. From the Scanning Electron Microscopy images of the tips and roughness measure of coverslip, we estimate the uncertainty in tip-substrate separation to be $\sim$ 2 nm. 

Another important point, which affect the accuracy of the tip-substrate separation is a possible relative tilt between them. As mentioned before, we utilize the voltage given for the scanner piezo to find out the gap. In our measurements, piezo voltage for zero of the separations is determined from the sudden change in QTF current amplitude when tip is about to make a contact with the substrate. If tilt exists between tip and substrate, one edge of the tip is closer to the substrate and may cause this sudden change in amplitude. For such a scenario, even when we measure zero separation there may exist a  small gap between the substrate and center of the tip. This is likely to provide an underestimate in  separation. To minimise the tilt, we take care to fix the tip perpendicular to the substrate and make  FIB cut on the tip which is  perpendicular to the axis of the fiber. Even with these cares, a slight tilt can be present. There is a way to check if  surfaces are exactly aligned parallel to each other while performing measurements. If there is a tilt, as the tip oscillate laterally, the vertical force fields are also sampled. This contribution is largely due to van der Waals force between the tip and the substrate and it is expected to be attractive. As shown by Kim et. al.\cite{kim_probing_2015}, it introduces a phase lag in QTF current. This phase lag is present when the relative tilt between tip and substrate is more than $0.3^0$. In all our measurements such phase lag is not seen. This amounts to a possible tilt of $0.3^0$ or less. Given the radius of our tip($\sim$ 400 nm) and a tilt of $0.3^0$, the gap size is likely to be about 2 nm  at the centre of the confinement region when one end of the tip is touching the substrate. Therefore, we consider a systematic error of 2 nm in all our measurements.  All the measurements of separations shown in our graphs are calculated from the piezo voltages and may have this possible systematic error.

Figure \ref{fig1} shows the representative autocorrelations obtained for Coumarin 343 diffusing through the gap between the tip and substrate held at three different separations. For separations of the order of 1 $\mu$m, the feedback control is turned off and the tip is pulled back from near-contact position by the distance mentioned in figure \ref{fig1}.

\section{Theory}
The average residential time spent by fluorophores in the observation volume determines the diffusion coefficient. It  is estimated from the autocorrelations of fluorescence intensity fluctuations. A similar experimental technique called Fluorescence Correlation Spectroscopy(FCS) deals with autocorrelation of fluorescence instensity from the confocal volume in bulk. Elson and Magde have derived an expression for  autocorrelation of intensities which accounts  for geometrical details of the observation volume, quantum yield of the fluorophore and the detection efficiency of the instrument \cite{magde_fluorescence_1974, thompson_fluorescence_1999}.
Fitting this expression to  experimental autocorrelation data yields the average residential time.  
However, this expression, which  is derived for a typical Fluorescence Correlation Spectroscopy (FCS) experiments needs modification in order to apply to our experimental data. In a typical FCS, the confocal volume is placed well inside the bulk of the solution and the movement of fluorophores are not restricted by any physical boundary. The illumination and collection is done using the same objective lens. We illuminate part of the confined region using the tip and emission is collected by an objective.

There are several  attempts to derive an expression to fit autocorrelation data from FCS when there are two parallel boundaries present in the confocal volume\cite{gennerich_fluorescence_2000}. For gaps much smaller than the height of confocal volume, this expression  can be approximated by the one for two-dimensional(2-D) diffusion\cite{gennerich_fluorescence_2000}. We use the expression describing 2-D diffusion in a typical FCS set up to fit our data and estimate the average residential time $\tau_D$.

For a freely diffusing fluorescent species, the FCS expression for autocorrelation is:

\begin{align}\label{eq:2dfcs_1species}
G(\tau)&=\frac{1}{<N>}\frac{1}{\left(1+\frac{\tau}{\tau_D}\right)}\left(1+\frac{T}{1-T}e^{- \tau / t_t} \right)
\end{align}

Here, $<N>$ is the average number of fluorescent molecules in the observation volume and $\tau_D$ is the residential time. "T" is the fraction of fluorophores in triplet state and $t_t$ is the triplet transition timescale. If there are two fluorescent species in the observation volume with different diffusion coefficients, the autocorrelation modifies as:

\begin{align}\label{eq:2dfcs_2species}
G(\tau)=\frac{1}{<N>} \left[\frac{\phi_1}{\left(1+\frac{\tau}{\tau_{D1}}\right)} +  \frac{\phi_2}{\left(1+\frac{\tau}{\tau_{D2}}\right)}\right]  \left(1+\frac{T}{1-T}e^{- \tau / t_t} \right)
\end{align}

Here, $\tau_{D1}$ and $\tau_{D2}$ are the residential time for species 1 and 2 respectively. And $\phi_1$ and $\phi_2$ are the corresponding weight factors for each species.
The weight factor for a particular fluorescent species depends on its fluorescence yield and fraction of that species in the average number of total fluorescent species present in the observation volume($<N>$). If $q_1$ and $q_2$ are the fluorescence yield of molecule 1 and 2, and "$f$" is the fraction of molecule 1. Then the weight factors; $\phi_1$ and $\phi_2$ are given by:

\begin{align}\label{eq:phi}
\phi_1&=\frac{f q_1^2}{\left[f q_1 + (1-f) q_2\right]^2}\\ \nonumber
\phi_2&=\frac{(1-f) q_2^2}{\left[f q_1 + (1-f) q_2\right]^2} 
\end{align}

In situations where the fluorescent molecule adsorbs or desorbs from the surface  along with diffusion. The FCS expression for autocorrelation is given by\cite{wirth_analytic_2001}:

\begin{align}\label{eq:2dfcs_adsorption}
G(\tau)&=\frac{1}{<N>} \left[\frac{\phi_1}{\left(1+\frac{\tau}{\tau_D}\right)}+\phi_2 e^{- k \tau}\right]\left(1+\frac{T}{1-T}e^{- \tau / t_t} \right)
\end{align}

The weight factors $\phi_1$ and $\phi_2$ are defined exactly like in equation \ref{eq:phi}. Instead of  subscripts 1 and 2 being different molecules, here 1 indicates the molecules undergoing diffusion and 2 indicates the faction of same molecules which are undergoing adsorption-desorption events. And "k" is the rate
constant of desorption. 

For typical FCS, the residential time is related to the radius of observation volume along lateral direction. This radius, $\omega_{xy}$, is the distance from the peak of Gaussian which determine the objective illumination to the $1/e^2$ value of the peak. This typically matched with the radius of confocal volume for FCS. The relation between residential time ($\tau_D$) and $\omega_{xy}$ is  
$\omega_{xy}^2 = 4 D \tau_D$
Here, "D" is the diffusion coefficient of the fluorescent molecule. This follows from Einstein's equation for mean square displacement in 2-D for the time period $\tau_D$.

Similarly, in our setup we can define the width of the observation volume with the fit value of $\tau_D$ from the 2-D FCS autocorrelation expressions. In case  where the tip illumination is broader than the width of confocal volume,  the width of observation volume is mainly determined by the width of confocal volume and is  fixed for a given tip and alignment. This implies that "D" is inversely proportional to $\tau_D$ for the measurements with same tip and same alignment.  The estimate of $\tau_D$ at  different separations with same tip and alignment, enables us to verify if diffusion is hindered due to nanoscale confinement. 

For a given tip and alignment we can get an expression:

\begin{align}\label{eq:D_ratio}
\frac{D}{D_{bulk}}=\frac{\tau_D^{bulk}}{\tau_D}
\end{align}

$\tau_D^{bulk}$ is the value measured at  large separations ($>$100 nm ) with the same tip, where the diffusion is bulk-like.

\section{Simulations} 

\begin{figure}
	\includegraphics[width=0.4\textwidth]{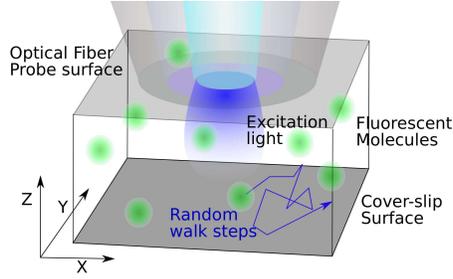}
	\caption{ A schematic to describe the Monte Carlo simulation box. \label{fig2}}
\end{figure}

The models described in previous section have assumptions which deviates from the experimental situation. Models uses 2D diffusion instead of confined diffusion and illumination from objective rather than from the tip. The quantification of diffusion coefficient may have systematic errors and hence its determination may not have significant confidence.  We can also estimate diffusion coefficients from the experimental  data using a Monte Carlo based fitting procedure\cite{ajith_new_2020}.

The schematic of the simulation box is given in figure \ref{fig2}. The top and bottom surfaces of the simulation box are the confining surfaces of tip and coverslip. Reflecting boundary conditions are used at these two surfaces. The separation between them is decided from the experiment. The other vertical sides of the simulation box have periodic boundary condition. The length and  breadth of the box is at least 3 $\mu $m each  to avoid the artifacts of periodic boundary conditions for a typical tip illumination of 1 $\mu$m diameter. Briefly, the simulation starts with a random distribution of fluorescent molecules with concentration used in the  experiment. The  fluorescent intensity obtained in confocal volume from this distribution of molecules under the tip illumination is calculated by taking an excitation profile, detection efficiency and other experimental parameters. In the next time step, the positions of fluorescent molecules are changed according to a certain diffusion coefficient. The fluorescent intensity is calculated again for the new distribution of fluorescent molecules. By performing such calculations for each time step,  a time trace of fluorescent intensity is obtained. The autocorrelation of the fluctuations in this time trace is calculated. 

As the values of concentration and fluorescence properties used in simulation and alignments in experiments can have slight variations from ideal expectations, the intercept of the simulated autocorretation may not match with the experimental autocorrelation. So before fitting the simulated autocorrelation with the experimental one, the intercept is fitted. This is done by multiplying a factor to the simulated autocorrelation to get a corrected one. The factor value is determined by finding the one which gives the least root mean square error between the simulated autocorreation and experimental autocorrelation. This way, corrected autocorrelations from the simulation for different diffusion  coefficients are obtained. Among these simulated autocorrelations the one which gives the minimum of root mean square error is the simulation fit. The diffusion coefficient corresponding to this fit is the diffusion coefficient of the fluorophore in the experiment. See ref. \cite{ajith_new_2020}  for more details about the simulations.

\section{Results and Discussion} 

\subsection{Diffusion in pure water}

\begin{figure}
	\includegraphics[width=0.4\textwidth]{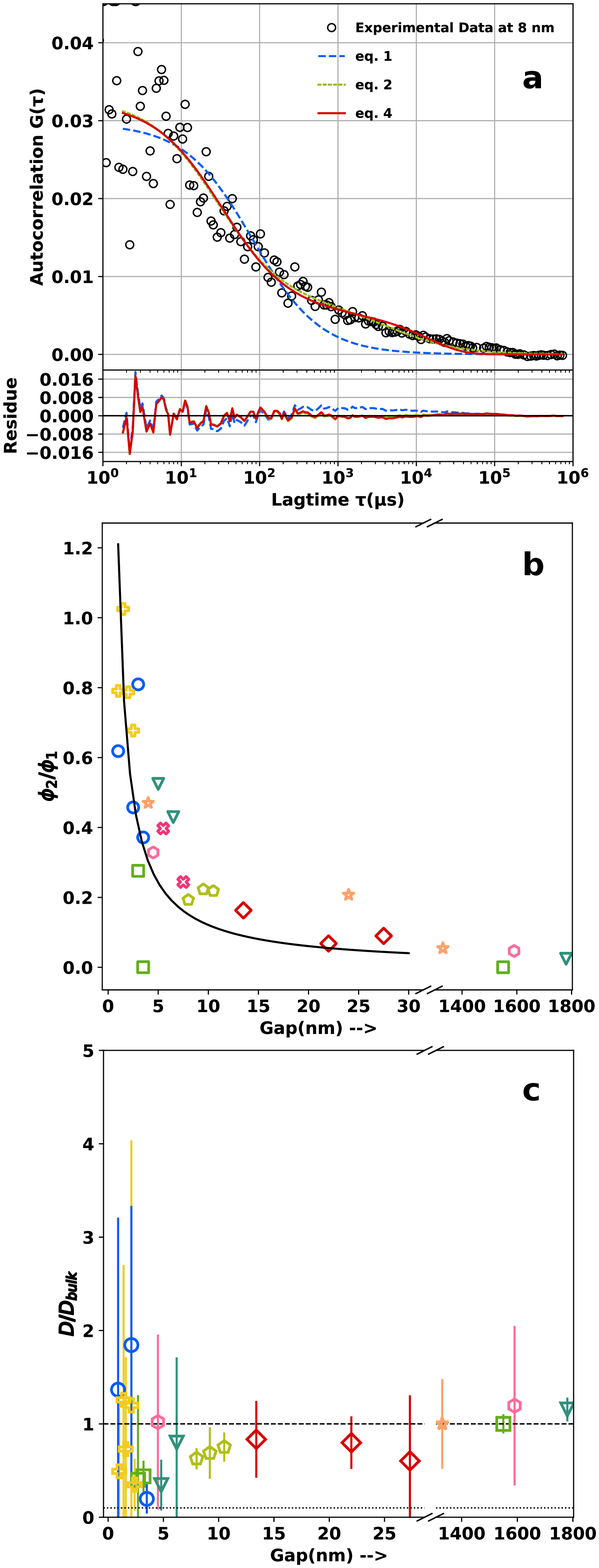}
	\caption{a) A representative curve for  autocorrelation of fluorescence  intensities from  a dye diffusing in 8 nm gap with best fits using equation 1 ( dashed blue line ),    equation 2 (dotted green line ) and  equation 4 ( continuous red line).  
		b) Ratio of weight factors of the adsorbing -desorbing fraction $\phi_2$ to diffusing fraction $\phi_1$. Solid line is the fit to $C_3/d$, where $C_3$ is a positive constant and d is separation.
		c) $D/D_{bulk}$  versus separation determined using equation \ref{eq:D_ratio}, with the faster time scale $\tau_D$ 
		\label{fig3}   }
\end{figure}

\begin{center} 
	\begin{table}[h!]
		\begin{tabular}{ | c || c | c | c |}
			\hline
			Parameters & Eq: \ref{eq:2dfcs_1species} & Eq: \ref{eq:2dfcs_2species} & Eq: \ref{eq:2dfcs_adsorption} \\ 
			\hline
			$1/<N>$ & 0.0296 & 0.0328 & 0.0207 \\  
			$\tau_D$ or $\tau_{D1}$ ($\mu$s) & 81 & 29 & 33  \\  
			$\tau_{D2}$ ($\mu$s) & - & 5037 & -  \\ 
			k$(s^{-1})$ & - & - & 68.5  \\ 
			$\phi_1$ & - & 0.81 & 1.312 \\ 
			$\phi_2$ & - & 0.19 & 0.252 \\
			T\textsuperscript{\emph{a}} & 0.6 & 0.0001 & 0.0001  \\
			$t_t(\mu s)$\textsuperscript{\emph{a}} & 0.002 & 10 & 0.03 \\
			\hline
			RMS residue & 0.003702 & 0.003123 & 0.003150 \\
			\hline
			
		\end{tabular}
		\caption{ The parameters after fitting equation 1, 2 and 4 to a representative experimental auto-correlation curve in Figure \ref{fig3}a. The slow component time scale $\tau_{D2}$ varies from 500 $\mu$s to 5000 $\mu$s for different  experiments. }
			\textsuperscript{\emph{a}} The part of autocorrelation which gets affected by triplet state transitions are typically below $\sim10 \mu s$ . Our data from confinement is very noisy below $\sim10 \mu s$. So the fit parameters values for T and $t_t$ will not be reliable. This doesn't affect the fit values of other parameters. 
			\label{table:1}
		
	\end{table}
\end{center}

Figure \ref{fig3}a shows a representative autocorrelation of Coumarin 343 diffusing in 8 nm gap with best fits of three different equations;  for single diffusing species (eq: \ref{eq:2dfcs_1species}), two diffusing  species (eq: \ref{eq:2dfcs_2species}) and one single diffusing species performing  adsorption-desorption events (eq: \ref{eq:2dfcs_adsorption}). Clearly, the single diffusing species poorly describes our data. The equation is not fitting well ( blue dashed line).
Earlier \cite{ajith_new_2020}, we have reported development of the instrument and presented preliminary data, wherein the autocorrelations below 100 $\mu$s were noisy and excluded from fit procedures. This led to a single timescale over which intensities were correlated and we concluded that there is slow-down in the diffusion under confinement.  We have now improved the instrument and the experimental protocol to  reduce the  noise in the early delay times. As a  result,  a larger range of data from 1 $\mu s$ to 1 s is now fitted.

The fit parameters are listed in the table \ref{table:1}.  Both, equation \ref{eq:2dfcs_2species} (green dotted line) and equation  \ref{eq:2dfcs_adsorption} (red continuous line) fit equally well. This means that both pictures are equally possible as far as quality of fit is concerned. 
However, since we are using a single fluorophore for these measurements, single diffusing species performing  adsorption-desorption events is more realistic model, provided the liquid under confinement is mostly homogeneous. In our experiments the confining walls are largely made of glass bearing negative charges. The large part of the tip is  made up of metal coating and may accumulate charges. While performing measurements it is difficult to avoid the tip-substrate contact owing to the occasional instability of the feedback loop. The tip also needs to make a momentary contact  during a approach-retract cycle. This cycle is essential for determining zero of the separation and dependence of tip-amplitude on the separation. Such intermittent tip-sample contact may smear the charges accumulated on the tip to  substrate and render the confining wall positively charged locally. This positive charge becomes an adsorption sites for negatively charged Coumarin 343. It is important to note that on rare occasions when tip does not make the substrate positively charged in this way, the slow component seen in the autocorrelation curve in \ref{fig3}a is not seen. This, however, happens rarely and is not reproduced in a controlled manner. In the next section we will discuss a remedy to mitigate this situation.

 The average number of molecules performing adsorption-desorption are same for different confinement gaps. This is because the area of interfaces in the observation volume from the tip and coverslip are the same for all the separation between coverslip and the tip. Thus,  the number of adsorption sites accessed by the fluorescent molecules for a given concentration will be the same and number of molecules undergoing adsorption-desorption events; $<N_2>=C_1$, a positive constant. The number of molecules undergoing diffusion should increase linearly with the gap as the observation volume increases linearly with increase in separation. So, $<N_1>= \pi r^2 d c= C_2 d$, where "r" is the radius of observation volume, c is the concentration of the fluorescent molecule in the confinement region. This is likely to be different from the bulk concentraion. $C_2= \pi r^2 c$ is another positive constant. Now we can calculate the fraction of diffusing molecules $f=<N_1>/<N>=<N_1>/(<N_1>+<N_2>)$. Substituting this "$f$" in the expressions for $\phi_1$ and $\phi_2$ and taking the ratio $\phi_2/\phi_1$, we can see that this ratio is inversely proportional to gap(d) with a positive proportionality constant "$C_3$". And $C_3=\frac{q_2^2 C_2}{q_1^2 C_1}$

\begin{align}\label{eq:phi_ratio}
\frac{\phi_2}{\phi_1}=\frac{C_3}{d}
\end{align}

Our data does follow this trend.  Figure \ref{fig3}b shows plot of the ratio of the weight factors of adsorbed to the diffusing fraction $\phi_2/\phi_1$ from  experiments at different tip-substrate separations. The values of $\phi_1$ and $\phi_2$ is obtained by fitting equation \ref{eq:2dfcs_adsorption} to the autocorrelation data. Each colour or type of symbol represents data from  experiments performed using different tips. The fit to $C_3/d$ confirms that there are fluorophores homogeneously distributed over the confinement region except the ones adsorbed over the tip and substrate. The average residential time $\tau_D$ of the diffusing species can then be estimated by fitting equation \ref{eq:2dfcs_adsorption} to all the data.

Using $\tau_D$  at different separations and $\tau_{bulk}$( fit to data at separations $\geq$ 100 nm),  we find $D/D_{bulk}$ as mentioned in equation \ref{eq:D_ratio}. Figure \ref{fig3}c shows $D/D_{bulk}$ values with-respect to gap. Each colour or type of symbol represents data from  experiments performed using different tips. The error values for this ratio are calculated from the standard deviation of the $\tau_D$ from the fitting procedure.  We find that $D/D_{bulk}$ fluctuates around the value 1 within 10 to 0.1. This means that the diffusion coefficient of Coumarin 343 in confined water, in gaps as low as 2 $\pm$ 2 nm, is of same order of magnitude to the it's bulk value.

\subsection{Diffusion in 1M salt solution} 

\begin{figure}
	\includegraphics[width=0.4\textwidth]{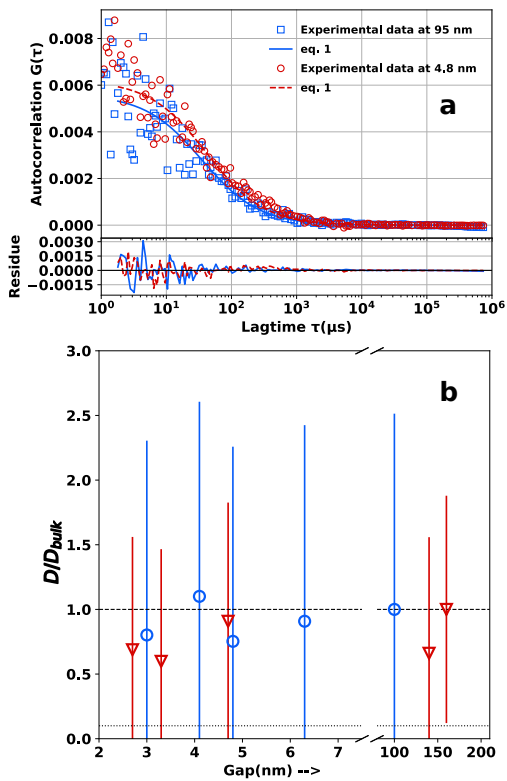}
	\caption{Representative autocorrelation curves  of Coumarin 343 in 1 M NaCl at  separations of 5 $\pm$ 2 nm and 95 $\pm$ 2 nm. The curves nearly overlap and do not have a component representing the adsorption-desorption  kinetics of the fluorophore to the glass surfaces. 
		The one  component description by equation (2) fits the data well at both separations. The overlap of both autocorrelations indicate that there is only fast moving bulk-like component.
		b)   $D/D_{bulk}$ vs separation with 1 M NaCl. The  horizontal lines are reference lines for D/ D$_{bulk}$ = 1 and 0.1 respectively.
		\label{fig4}  }
\end{figure}

We further confirm our findings by mitigating the  issue of fluorophore adsorption in a separate experiment. We performed experiments by adding  1M NaCl to the solution. Adding salt  is known to reduce the charge mediated adsorptions by screening the local adsorption sites with counter-ions(ions of opposite charge compared to the adsorption site)\cite{daniels_dye_2010}.  We choose to add 1M NaCl as similar amount  is found to enhance the probability of dynamic solidification of water\cite{khan_nacl-dependent_2016}. Moreover, the Debye length at this concentration is 0.3 nm \cite{schoch_transport_2008}.  Representative autocorrelation curves for Coumarin 343 in confined 1M NaCl solution are shown in figure \ref{fig4}a and \ref{fig5}a for gaps 2.6 $\pm$ 2,  4.8 $\pm$ 2 and 95 $\pm$ 2 nm. The equation for single  diffusing species with no adsorption (eq. \ref{eq:2dfcs_1species}) fits well on all experimental autocorrelations.  We deduce that  adsorptions are negligible and the  autocorrelation curve is mainly determined by diffusion. The $\tau_D$ values corresponding to these representative autocorrelations for 2.6 $\pm$ 2,  4.8 $\pm$ 2 and 95 $\pm$ 2 nm were 46, 52, 39 $\mu s$ respectively. 

The $\tau_D$ values from the fit are used to calculate $D/D_{bulk}$. The $D/D_{bulk}$ values are plotted against the confinement gap in figure \ref{fig4}b. The error value for each $D/D_{bulk}$ is calculated from the standard deviation of $\tau_D$ from the fitting procedure. Similar to pure water, the $D/D_{bulk}$ values of Coumarin 343 fluctuates around 1 between 10 and 0.1.  The diffusion coefficient of Coumarin 343 in confined 1 M NaCl solution, for confinements as low as $\sim$ 2.6 $\pm$ 2 nm, is in the same order of magnitude of it's bulk value.

\subsection{Monte Carlo to estimate Diffusion}

\begin{figure}
	\includegraphics[width=0.4\textwidth]{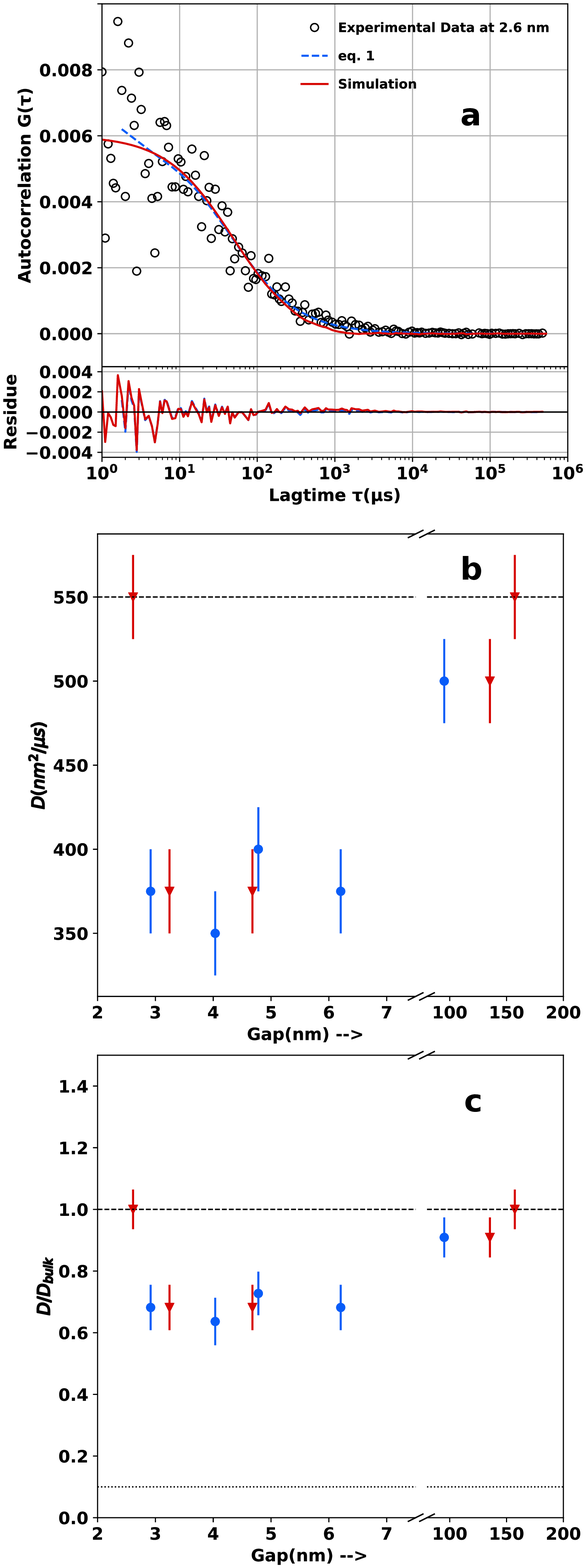}
	\caption{a) The fit of  simulated data (red continuous line) to autocorrelation curve  (black symbols)  for confinement with tip-substrate separation of  2.6 $\pm$ 2 nm. The salt concentration is 1 M. Monte Carlo simulations are run with different  diffusion coefficients till simulations and experiments fit well.   The fit yields a diffusion  coefficient of 550$\pm$25 $nm^2/{\mu s}$ b) Diffusion coefficient D  versus separation. D is  estimated by fitting simulations to the experiments,  c) D/D$_{bulk}$ versus separation. The  horizontal lines are reference lines for D/ D$_{bulk}$ = 1 and 0.1 respectively.  
		\label{fig5} }
\end{figure}

The data from NaCl solution are further analysed with the Monte Carlo fitting procedure. A typical simulation fit to the autocorrelation from nanoconfined NaCl solution is given in figure \ref{fig5}a. This data is for Coumarin 343 diffusing in  confinement gap of 2.6 $\pm$ 2 nm having 1 M NaCl solution. 
The diffusion coefficient corresponding to the auto-correlation data from the experiments is estimated from this fitting. For this particular data simulated autocorrelation for D=550 $nm^2/{\mu s}$ gave the best fit. The figure \ref{fig5}b shows the fitted diffusion coefficient for different separations in two different experiments  with different tips. The error bars for each diffusion coefficient is the minimum change in diffusion coefficient around the fitted value such that a significant change($> 10^{-6} $) in root mean square error is obtained. We  also find the $D/D_{bulk}$ values. The figure \ref{fig5}c shows the $D/D_{bulk}$ with respect to confinement gap.

\subsection {Discussion}

The autocorrelation of fluorescence intensity fluctuations revealed that there are two  timescales over which intensity is correlated for the Coumarin 343 molecules diffusing in nanoconfined water.  It is found that fast time scale corresponds to the average time  required to cross the  width of observation volume,  provided  the molecule has diffusion coefficient similar to its bulk value. The  slower time scale disappeared in measurements with 1 M NaCl . Adding NaCl is known to avoid adsorption of solute molecules over substrates due to screening of electrostatic interactions. The Coumarin 343 molecule(11-oxo-2,3,5,6,7,11-Hexahydro-1H-pyrano[2,3-f]pyrido[3,2,1-ij]quinoline-10-carboxylic acid;\ch{C16H15NO4}) has the caboxylic-acid group. It gains negative charge in pH $\sim$7. One of the confining surfaces, the cover-slip is cleaned with UV-Ozone before the  experiments which enriches the surface with hydroxyl-terminations. This cover-slip in pH $\sim$7 water get negative charges\cite{parks_isoelectric_1965}. As discussed earlier,  the second confining surface, the tip-surface may accumulate unknown static charges owing to metal coating. This may render the glass surface positively charged. By adding NaCl to pure water, we  introduce co-ions and counter-ions. The counter-ions get attracted to the local charge centres on the surfaces and forms Electrical Double Layer(EDL). The Debye length associated for this EDL in 1 M NaCl is $\sim$0.3 nm\cite{schoch_transport_2008}. Thus  $\sim$0.3 nm EDL is present over both of the confining surfaces. The Coumarin 343 diffusing in this confined gap experience highly shielded electrostatic potential due to EDL from the surfaces. This helps in avoiding the adsorption of  Coumarin 343 over the confining surfaces. The disappearance of slower time scale in autocorrelation for 1 M NaCl confirmed  that adsorption-desorption process is the reason for slower time scale in nanoconfined water without salt.

Furthermore, the ratio of weight factors of the fraction undergoing adsorption-desorption process and the fraction of molecules executing diffusion($\phi_2/\phi_1$) show a trend with respect to confinement gap. This is consistent with our physical picture that number of molecules undergoing diffusion will increase linearly as the gap is increased while the number of molecules undergoing adsorption-desorption will remain unchanged. 
By fitting such a model,  we could extract the residential time scale due to diffusion for various separations. Further, we could estimate the ratio of diffusion coefficient with it’s bulk value.  We find that the diffusion coefficient of the tracer molecule is similar to the bulk value even for confinements as small as 2 $\pm$ 2 nm. Similar behaviour is also seen in 1 M NaCl solution confined till 2.6 $\pm$ 2 nm. This behaviour is independent of analysis method used to extract the diffusion coefficient. The estimate of diffusion coefficients at all separations using Monte Carlo simulation based fitting procedure, yields the same result.

In figure \ref{fig1}, we see that the y-intercept of autocorrelation from different gap are different. From  equation \ref{eq:2dfcs_adsorption} we realise that the intercept is given by $(\phi_1+\phi_2)/(<N>(1-T))$. And $\phi_1$ and $\phi_2$ are defined according to equation 3.  This behaviour of intercept in figure \ref{fig1} is not simple to explain because of the nonlinear dependence of $\phi_1$ or $\phi_2$ with the fluorescence yield and fraction of diffusing molecules. With a systematic study of this variation of intercept with gap, one could estimate the fluorescence properties of adsorbed and unadsorbed molecules. This is outside the scope of this paper.

We discuss the uncertainty estimates in determining the slow and fast time-scales. This is not possible by performing a statistical analysis presented by Wohland et al.\cite{wohland_standard_2001} for a typical FCS experiment. This requires the experiments to be performed multiple number of times at the same separation and experimental conditions. Such analysis is possible in a typical FCS experiment, wherein the observation volume is in the bulk of liquid and any number repetitions  under the same conditions are possible. Although, the tip-sample separation can be held fixed over several seconds or minutes in our experiments, the same is not true about the objective and the cover-slip. Thus,  it is difficult to estimate uncertainty in  diffusion coefficient under confinement using a statistical analysis used in a typical FCS experiments. However, the errors in diffusion coefficient are determined through fitting procedures and Monte Carlo method.

The interfacial water has  unique properties compared to bulk water.  Yet, it is not clear under what length scales of confinements the interfacial properties become prominent.  This may even depend on the physical property being measured. For instance,  the out-of-plane dielectric coefficient of nanoconfined water over hydrophobic confining surfaces was measured to be   $\sim$2, which is much smaller than the bulk value $\sim$80\cite{fumagalli_anomalously_2018}. This so called electrically dead nanoconfined water is thought to be due to the reduction of rotational freedom of water molecules near interface\cite{fumagalli_anomalously_2018,sato_hydrophobic_2018}. The capacitance measurements revealed that a gap size of more than 100 nm is required for water to have a bulk-like dielectric coefficient\cite{fumagalli_anomalously_2018}. However, behavior of nanoconfined water, when one measures  translational diffusion seems different. There are reports of many orders of change in viscosity for interfacial water\cite{zhu_viscosity_2001,li_nonlinear_2008}. This means it may  exhibit a slow-down in self-diffusion  as well as reduced tracer diffusion. The self-diffusion of water is probed in ensemble measurements using NMR or QENS\cite{walther_hansen_self-diffusion_1995,bellissent-funel_single-particle_1995,mitra_dynamics_2001,takahara_neutron_2005}. These measurements find that, at  room temperature the translational diffusion of water is similar to  bulk value within an order for confinements as small as $\sim$ 2 nm. However, below this  separations the molecular effects,  such as restriction in dye orientation and rotational diffusion are expected to affect the diffusion of the dye. In our measurements the diffusion coefficient retaining a bulk-like value  at or below   2 nm  could be owing to the limitations in determining the gap size accurately. From our data and analysis we can conclude that diffusion is certainly bulk-like at separations of  5 nm and above. Figure \ref{fig3}c, \ref{fig4}b, and figure \ref{fig5}c show that the diffusion coefficient of Coumarin 343 in nanoconfined water for gaps as small as 5 nm is within the same order of magnitude of the bulk value. This conclusion is consistent with NMR and QENS studies.  However, like dielectric coefficient of interfacial water, the diffusion of tracer molecules can be anisotropic\cite{schlaich_water_2016}. We measure the diffusion of fluorescent tracer molecules on an average moving parallel to the confining surfaces. The local out-of-plane diffusion might be different from what we measured. 

The majority of rheological measurements with AFM or SFA have shown a  large  deviation of viscosity only below 1 nm of hydrophilic confinements\cite{li_nonlinear_2008,zhu_viscosity_2001,sakuma_viscosity_2006}. Above 2 nm the average viscosity or dissipation measured were similar to bulk values. Thus, the AFM and SFA measurements also imply an average bulk-like diffusion for nanoconfined water in gaps more than 2 nm.  Molecular dynamics simulations wherein, the in-plane transnational diffusion timescale for water in 2.44 and 1.65 nm gaps are found to be similar to bulk value\cite{leng_fluidity_2005},  also agree with our conclusion.

There are  few reports of measurement of translational diffusion of tracer dye in water confined at nanoscale. The measurement with varying gap size below 10 nm is not routine.   For sub-micrometer confinements many particle tracking measurements are done using colloidal tracers\cite{faucheux_confined_1994, carbajal-tinoco_asymmetry_2007}. These measurements are mostly consistent with fluid dynamic description of motion of a particle near interface. The smallest possible  confinement is limited by the size of the colloidal tracer particle. The choice of small fluorescent molecules as tracer allows the confinement to be as small as  1-10 nm.   Zhong et al. measured diffusion of Rhodamine B in ethanol taken inside nanochannels of height 8 nm and found a reduction in diffusion coefficient by 2 orders\cite{zhong_fluorescence_2018}. Kievsky et al. measured diffusion of Rhodamine 6G in water in porous material with pore diameter of 2.9 nm. They found an eight order reduction in diffusion coefficient\cite{kievsky_dynamics_2008}. Santo et al. measured diffusion of Rhodamine 6G in water in nanochannels of height 30, 20 and 10 nm. They found the diffusion coefficient to be similar to the bulk value in 30 and 20 nm channels and a 25\% slow-down is seen in 10 nm channel\cite{de_santo_subdiffusive_2010}. Grattoni et al. measured diffusion of Rhodamine B in water in nanochannels having height 200 and  1 $\mu$m, as well as 13 and 5.7 nm\cite{grattoni_device_2011}. They found a reduction of diffusion coefficient as the channel height is reduced. The lowest diffusion coefficient they measured (in 5.7 nm channel) is still within an order of the bulk value.  Our measurements on diffusion of Coumarin 343 agrees well with Santo et al. and  Grattoni et al.

The reasons for the drastic slow-down in the studies by Zhong et al. and Kievsky et al. is not clear. Usually this slow-down is attributed to the geometric confinement effects, surface interactions and possible viscosity difference of nanoconfined water. It is clear from our findings that the  diffusion appears slowed down  due to  adsorption-desorption.  Then the apparent diffusion coefficient calculated from the slower residential timescale due to adsorption will be few order less even for our data\cite{ajith_new_2020}.  It is possible that  the  slow-down reported  by Zhong et al. and Kievsky et al. is due to adsorption-desorption interactions. 

In order to compare our measurements better with existing literature we also performed  measurements with Rhodamine 6G. The cationic Rhodamine 6G is known to adsorb over glass surfaces\cite{chen_fluorescence_2008}. This adsorption is due to electrostatic and hydrophobic interactions\cite{mubarekyan_characterization_1998}. We found that, while we could avoid the mild adsorptions of Coumarin 343 by adding NaCl,  Rhodamine 6G have strong adsorption even in 1 M NaCl. The autocorrelation we measured  for Rhodamine 6G have large adsorption-desorption contribution even for 1 $\mu$m separations. As the  contribution of diffusing species to the  autocorrelation is less dominant, the average residential  time estimated from  fitting procedure is unreliable.  This unreliability in fitting parameters from the adsorption-desorption FCS equation when adsorption is dominant has been pointed out before\cite{wirth_analytic_2001}. In the work of Kievsky et. al., manipulating the pH or using organic solvent did not change the eight order reduction in diffusion coefficient. We also faced similar difficulties with Rhodamine 6G. However, measurements performed using Alexa Fluor 568 were successful and supports conclusions drawn using Coumarin 343. We observe that the choice of fluorescent molecules and interfaces  matters in these type of experiments.

The apparent diffusion coefficient which includes adsorption-desorption interactions can vary with separation even though the diffusion coefficient and timescale for adsorption-desorption interactions(1/k) are same for all separations. This is because the fraction of diffusing population changes with separations as we had seen in the trend of $\phi_2/\phi_1$ with gap. The time scale for this apparent diffusion can be estimated from the residential time scale for diffusing molecules and the residential time scale of the molecules undergoing adsorption-desorption. The timescales $\tau_{D1}$ and $\tau_{D2}$ from equation \ref{eq:2dfcs_2species} can be interpreted as these timescales respectively. If we can extract out the values of $f$, the fraction of diffusing molecules, from $\phi_1$ or $\phi_2$, the residential timescale of apparent diffusion coefficient($\tau_{apparent}$) can be estimated. It is the weighted average of $\tau_{D1}$ and $\tau_{D2}$, given by $f\tau_{D1}+(1-f)\tau_{D2}$. For estimating $f$ from $\phi_1$ or $\phi_2$ one needs to know the fluorescence yields of the fluorescent molecule away($q_1$) and near to the surfaces($q_2$) independently.  Qualitatively, we can see that for large separations we have $f$ approaching 1 and $\tau_{apparent}$ will approach to bulk-like diffusion timescale, $\tau_{D1}$. And for small separations, $\tau_{apparent}$ approaches the slow residential time scale due to adsorption-desorption. For instance, if we consider $\tau_{D2}$ value shown in table \ref{table:1} for a sub-10 nm confinement, it is two orders larger than the bulk diffusion timescale. Thus we expect two order slower apparent diffusion when adsorption-desorption interaction is included.
Our goal in this paper is to understand the dynamics in the confinement which is not driven by the surface effects but is a result of confinement itself. We conclusively show that there is no alteration in diffusion coefficient due to confinement alone. The slow-down of 2 orders seen in our experiments in the absence of salt,  which is dominated by the measurement of apparent diffusion,  can actually be attributed to the adsorption effects.

The length scale below 100 nm is particularly important in nanofluidics\cite{eijkel_nanofluidics_2005,bocquet_nanofluidics_2010,sparreboom_principles_2009}. Nanofluidics is applied in fields of iontronics\cite{karnik_electrostatic_2005,yan_nanofluidic_2009}, sequencing of biopolymers\cite{marie_integrated_2013,chinappi_protein_2018}, and energy harvesting\cite{ren_slip-enhanced_2008}. In this field,  the water at nanoscale is considered  to have bulk-like viscosity in channels above 1 nm\cite{bocquet_nanofluidics_2010}. Our conclusions,   which are arrived at   from independent diffusion measurements,  support this treatment of water at nanoscale. Usage of fluorescence correlation based measurements like Fluorescence Correlation Spectroscopy(FCS) is rare in nanofluidics.  As FCS has been used to study reaction kinetics\cite{thompson_fluorescence_1999,magde_fluorescence_1974}, flow and diffusion measurements\cite{thompson_fluorescence_1999,magde_fluorescence_1974}, extending it to nanofluidics will open-up many new possibilities. Our study shows how to interpret the timescales in fluorescence autocorrelation. This is beneficial for fluorescence correlation based measurements in nanofluidics and close to an interfaces.\\

\section {Conclusion}
In conclusion, we measured  diffusion of  tracer dye  in water confined to nanoscale, where the separation between the confining walls is as small as $\sim$ 5 nm. We found that the the diffusion coefficient of the tracer dye remains unchanged for such small gaps. It indicates that  viscosity of nanoconfined  water is bulk-like for separations of $\sim$ 5 nm. These conclusions are consistent with molecular dynamics simulations of water confined to similar gaps and many reports of self-diffusion of nanoconfined water using quasi-elastic neutron scattering experiments. 

\section{Author's contributions}
SP designed  the instrument and planned experiment along with VJ. VJ built the instrument,   performed  experiments  and did data analysis. VJ and SP wrote the manuscript together.

	\begin{acknowledgement}
		
		VJ acknowledges fellowship from DST- INSPIRE, Govt. of India. The work is supported by  intramural grant from IISER Pune 
		
	\end{acknowledgement}
	
%
%
	
	\bibliography{achemso-demo}
	
\end{document}